\documentclass[12pt,preprint]{aastex}
\usepackage{graphics}
\slugcomment{} \shorttitle{SDSS J104807.74+005543.5: Binary
Quasars?} \shortauthors{ZHOU ET AL.}

\received{2003 November 29}
\begin{document}

\title{Obscured Binary Quasar Cores in SDSS J104807.74+005543.5?}
\author{Hongyan Zhou, Tinggui Wang, Xueguang Zhang, Xiaobo Dong, and Cheng Li}
\email{mtzhou@ustc.edu.cn}

\affil{Center for Astrophysics, University of Science and Technology of
China, Hefei, 230026, China}

\begin{abstract}
We report the discovery of a possible close binary system of
quasars in SDSS J1048+0055. The [OIII]$\lambda\lambda$4959,5007
emission lines are clearly double-peaked, and two discrete radio
sources with a projected physical separation of $\sim 20$ pc are
found in the VLBA milliarcsec resolution image at 8.4 GHz. Each of
the [O III]$\lambda \lambda 4959,5007$ doublets and H$\beta$ can
be well modelled by two Gaussians and the line ratio, $[O
III]\lambda 5007 /H\beta \sim 7$, is typical of Seyfert 2
galaxies. No broad component of H$\beta$ was detected and its $[O
III]\lambda 5007$ luminosity, $L_{[OIII]}\approx 9.2\times
10^{42}$ erg $s^{-1}$, is comparable to luminous quasars and is a
few ten times more luminous than typical Seyfert galaxies. One
natural interpretation is that SDSS J1048+0055 contains two close
quasar-like nuclei and the BLR around them are obscured. Other
possible models are also discussed. We suggest that double-peaked
narrow emission line profile may be an effective way of selecting
candidates of binary black holes with intermediate separation.
\end{abstract}


\section{Introduction}

Black holes are the simplest objects in the universe and merging
of Binary Black Holes (BBH) can produce strong gravitational
radiation. Hence, BBH systems are ideal laboratories for exploring
and testing strong field relativistic physics. There is a growing
body of evidence that black holes exist on many mass scales, from
stellar mass, up to $\sim $ a few $10^{9}$$M_{\odot }$. Numerous
clues, especially those provided by the Hubble Space Telescope,
suggest that almost every galaxy harbors a Supermassive Black Hole
(SMBH) in its center (e.g., Sarzi et al. 2001; Pinkney et al.
2003). According to the hierarchical merger scenario of galaxy
evolution, coalescence is one of the most important processes in
galaxy evolution and a typical galaxy may experience between one
and several such episodes during its lifetime (Struck 1999).
Merger of galaxies can bring two SMBHs together eventually to form
a BBH, which can survive for about one Hubble time in spherical
galaxies, axisymmetric galaxies or certain triaxial galaxies (Yu
2002). Therefore the presence of two SMBHs in one galaxy forming
supermassive BBH, should not be a rare phenomenon.

Enormous efforts have been made in recent years to find
observational evidence for the existence of supermassive BBH.
Evidence is sought in two directions: spatially resolved systems
where the BBH can be identified morphologically and unresolved
systems where the existence of the BBH can be inferred in various
ways (see Komossa 2003 for an extensive review). Examples of the
latter are as follows. Winged or X-shaped radio sources were
suggested to be BBH merger remnants (e.g., Merritt \& Ekers 2002;
Wang et al. 2003; Liu, Wu, \& Cao 2003). Helically distorted radio
jets observed in some objects were interpreted sometimes as due to
the orbital motion or precession of the black hole producing the
jets (e.g., Sudou et al. 2003). Some BBH models predict the
surface brightness of the host galaxies would decrease toward the
galactic center and such an effect has indeed been observed in a
few early-type galaxies (e.g. Lauer et al. 2002). Rather more
direct evidence comes from resolved systems: several pairs of
quasars with almost the same redshift and angular separations
between $\sim 3-10"$ have been observed and were taken to be real
binary quasars. The projected physical distance between these
potential binary quasars is usually a few ten kpc (Mortlock,
Webster, \& Francis 1999). NGC 6240 is perhaps the only case where
binary active nuclei are separately identified: the high
resolution observation of Chandra revealed two hard X-ray nuclei
in this interesting object and the projected physical separation
of about 1.4 kpc (Komossa et al. 2003), which is the smallest
among known binary AGN. As far as we know, no quasar pair with
physical separation $\lesssim 1$ kpc has been yet found.

Of the above lines of evidence for the existence of supermassive
BBH, binary AGN is of particular interest. It must, however, be
admitted that apart from NGC 6240, the evidence so far cannot be
considered as conclusive and efforts must be continued to seek out
suppermassive BBH. In this letter, we report the discovery of a
new binary quasar candidate, SDSS J104807.74+005543.5 (hereafter
SDSS J1048+0055). It is identified from the Sloan Digital Sky
Survey Data Release 1 (SDSS DR1, Abazajian et al. 2003) among
$\sim 100$ low redshift galaxies that show double-peaked narrow
emission lines, most of which are active. Apart from the presence
of such lines, the other piece of evidence that sustains the BBH
interpretation is that its high angular resolution VLBA image
reveals two discrete radio sources, possibly corresponding to two
different active nuclei. The [OIII] luminosity of the two nuclei
are in the range for quasars. The discovery of SDSS J1048+0055 and
more than one hundred of double-peaked narrow line objects in the
spectral dataset of the SDSS DR1 might suggest that this be an
effective way of selecting candidates of binary black holes with
intermediate, sub-kpc separations. We postpone detailed analyse of
the whole sample of double-peaked narrow emission lines to a
forthcoming paper. Throughout this letter, an $H_{0}=$70 km~s$^{-1}$~Mpc$%
^{-1}$, $\Omega_{M}=0.3$, and $\Omega_{\Lambda} =0.7$ cosmology is
assumed.

\section{Data analysis}

\subsection{Optical spectrum}

SDSS J1048+0055 is observed by the SDSS only because it is the
counterpart of a FIRST radio source and classified as 'QSO' by the
spectroscopic pipeline of the SDSS (Stoughton et al. 2002). %Its
%optical spectrum in the observer's frame is shown in Figure 1 with
%recognizable emission lines labelled. 
It falls in the DR1 quasar
catalog (Schneider et al. 2003) due to its high luminosity
($M_{i}=-22^{m}.63$) and seemingly broad line width.

The optical continuum of SDSS J1048+0055, which is rather noisy,
is fitted with a power law. The fitted continuum is subtracted
before measurement of the emission lines. Because the
[OIII]$\lambda \lambda $4959, 5007 doublets clearly show a
double-peaked profile and a rather broad wing, the continuum 
subtracted spectrum in the H$\beta 
+$ [OIII]$\lambda \lambda $4959, 5007 regime is fitted with $2\times
3+1$ Gaussian components, i.e., two sets of three Gaussians for
the two peaks and the broad wing of the [OIII]$\lambda \lambda
$4959,5007 doublets and one Gaussian for H$\beta $. Within each
set of the components for [OIII], the centroids of the three
Gaussians are forced to correspond to a same redshift; their
widths to a same Doppler broadening; also the intensity ratio of
[OIII]$\lambda $5007 to [OIII]$\lambda $4959 is fixed at its
theoretical value. The final fit is done through minimization of
$\chi ^{2}$ and the result is shown in Figure 1 upper panel. The
lower panel displays the [OII]$\lambda $3727 line, which is the
2nd strongest after [OIII]$\lambda $5007. However, the [OII] line
is single-peaked and can be satisfactorily modelled by a single
Gaussian.

We would like to point out that H$\beta$ might also be double-peaked. 
When fitted with a single Gaussian, its line width $FWHM(H\beta )\approx 
1200 km s^{-1}$ is much broader than [OII] and each of the [OIII] peaks. 
The line can be fitted, comparably, with the template of [OIII] line 
profile,  Other high
ionization narrow lines, such as [NeIII]$\lambda $3869 and [NeIII]
$\lambda $3967 also seem to show double-peaked profile but higher
quality spectral observation is needed to confirm this feature. We
did not analyse other emission lines because the S/N ratio is not
high enough for us to draw any significant conclusion. The
measured line parameters are assembled in Table 1.
\clearpage
\tablenum{1}
\begin{deluxetable}{lllll}
\tablecaption{Line Parameters of SDSS J1048+0055.\label{tbl-1}}
%\rotate
\tablewidth{0pt} \tablehead{
\colhead{Line\tablenotemark{a}} & \colhead{Center Wavelength\tablenotemark{b}} & redshift & \colhead{Flux\tablenotemark{b}} & \colhead{FWHM}  \\
\colhead{} & \colhead{$\mathring{A}$} & & \colhead{10$^{-17}$erg
s$^{-1}$ cm$^{-2}$} & \colhead{km s$^{-1}$} } \startdata
$[OIII]\lambda 5007_{r}$ & $8238.5\pm 0.3$ & 0.6450 & $161\pm 11$ & $553\pm 25$   \\
$[OIII]\lambda $5007$_{b}$ & $8218.5\pm 0.5$ & 0.6410 & $209\pm 13$ & $610\pm 24$   \\
$[OIII]\lambda $5007$_{br}$ & $8224.7\pm 0.9$ & 0.6423 & $151\pm 11$ & $2573\pm 198$   \\
$[OIII]\lambda $4959$_{r}$ & 8159.5 & 0.6450 & 52 & 553   \\
$[OIII]\lambda $4959$_{b}$ & 8139.7 & 0.6410 & 67 & 610   \\
$[OIII]\lambda $4959$_{br}$ & 8142.5 & 0.6423 & 49 & 2573   \\
$H\beta $ & $7987.5\pm 1.0$ & 0.6422 & $76\pm 6$ & $1238\pm 102$   \\
$[OII]\lambda $3727 & $6122.5\pm 0.2$ & 0.6422 & $135\pm 9$ & $721\pm 36$ \\
$MgII\lambda $2800 & $4600.1\pm 1.0$ & 0.6430 & $98\pm 10$ &
$1806\pm 141$

\enddata
\tablenotetext{a}{'r', 'b' and 'br' denote red, blue and broad
component of the $[OIII]\lambda \lambda 4959,5007$ emission lines,
respectively. } \tablenotetext{b}{The central wavelengths are in the 
observer's frame.}
 %and the line widths (FWHM, full width at half-maximum 
%intensity) are values in the rest frame. }

\end{deluxetable}
\clearpage
\subsection{Radio properties}
SDSS J1048+0055 was observed many times in several radio
band-passes, from 0.365 to 8.4 GHz. It was observed at 1.4 GHz
during NVSS in 1995-02 (Condon et al. 1998) and during FIRST in
1998-08 (White et al. 1997). There is no significant difference in
the flux observed on these two occasions ($S_{\nu,NVSS}=270.9\pm
8.1 $ mJy and $S_{\nu,FIRST}=270.7$ mJy). This flux is also
marginally consistent with an early measurement in the Green Bank
1.4 GHz Northern Sky Survey, which is $S_{\nu,1.4 GHz}\approx 299$
with uncertainty of $\sim 25-30$ mJy (White \& Becker 1992).

However, the 4.85 GHz flux of $340\pm 21$ mJy given in the
MIT-Green Bank 5 GHz Survey Catalog (MGBS, Bennett et al. 1986) is
different (at $\sim 2\sigma $ level) from that presented by
Gregory \& Condon 1991 ($266\pm 37$ mJy), Becker, White, \&
Edwards 1991 ($262\pm 39$ mJy) and Griffith et al. 1995 ($262\pm
39$ mJy). The flux difference between the MGBS and the later
surveys cannot be explained by the larger beamsize of MGBS
(FWHP$=2'.8$): 1) because the source is compact, and 2) because
within $5'$ of the source NVSS detected no other sources while
FIRST only detected two with insignificant fluxes around 1 mJy. It
seems likely that this variation with an amplitude $\sim 39\% $ is
real. This interpretation is, moreover, consistent with its
compact morphology and its flat spectrum.

The overall radio spectrum is moderately flat with a spectral
slope of $\alpha _{\nu ,0.365-8.4GHz}=0.13\pm 0.03$ (S$_{\nu
}\varpropto \nu ^{-\alpha }$). The radio luminosity between
$0.365-8.4$ GHz, $ L_{0.365-8.4GHz}\approx 4.2\times 10^{43}$ erg
s$^{-1}$, is quite high even for a radio-loud quasar. The flatness
of the spectrum is consistent with the compact radio morphology.
It is unresolved by FIRST at angular resolution $\sim$ 5''.4. This
object is also observed by the VLBA Calibrator Survey (VCS1,
Beasley et al. 2002) and remains unresolved at 2.3 GHz with
resolution ellipse of $\sim 5\times10$ milliarcsecs. However, the
higher resolution 8.4 GHz radio image ($\sim 1.5\times 3$
milliarcsecs) clearly shows two compact sources with the west
source $\sim$ 30 times stronger than the east one (Figure 2). The
angular offset between the two sources is about 2.5 milliarcsecs,
corresponding to a projected physical distance of $\sim 20$ pc at
a redshift of 0.6422 derived from the [OII] and $H\beta $ emission
line. The large difference in the radio flux, the small angular
separation and flat spectrum conspire to indicate that we are dealing 
with possible two separate radio sources, rather than with two lobes 
of a single source. Further measurement of the radio spectrum of the 
weak component would be necessary to confirm this.

\section{Discussion and future prospect}

\subsection{SDSS J1048+0055 as obscured binary quasars}

A straightforward interpretation of the double-peaked profile of [OIII]$%
\lambda \lambda $4959,5007 is that they come from two distinct narrow
emission line regions (NLR) around two active nuclei. The [OIII]$%
\lambda $5007 luminosity of the red and blue peaks,
$L_{[OIII],red}\approx 2.8\times 10^{42}$ erg s$^{-1}$ and
$L_{[OIII],blue}\approx 3.6\times 10^{42} $ erg s$^{-1}$, is
comparable to that of luminous quasars in the BQS sample, and at
least 10 times more luminous than Seyfert galaxies. SDSS
J1048+0055 is rather faint in the optical in comparison with the
radio. The flux ratio of radio to optical or the Radio-Loudness,
$RL\equiv f_{r}/f_{o}\sim 1.5-2.1\times 10^{4}$, is
extraordinarily large. Such a large RL implies that the nuclear
emission in the optical is heavily obscured. The radio morphology
also suggests that two active nuclei may be present in the center
of SDSS J1048+0055 as was pointed out in the last section. 
If the two discrete radio sources revealed by VLBA proved to be
stationary and if the weak component also shows flat spectrum  by 
further VLBI observations, thereby thus confirming
their BBH status, then we can say that the separation between the
BBH should not be much larger than tens of pc. Then SDSS J1048+0055 
would be the only galaxy harboring BBH with such a small separation 
and the system may be at the last stage before the two nuclei (BHs) become 
a bound system (Yu 2002). The fact that [OIII]$\lambda \lambda 4959,5007$  
show double-peaked profile while [OII]$\lambda 3727$ does not is 
understandable because high ionization emission line regions are often 
more compact than low ionization ones. It is likely that double-peaked 
[OIII] emission line is mainly concentrated in the two cores of NLR 
whose centers coincide with the two radio peaks while the narrower and single-peak 
[OII] may be emitted by the whole NLR. Therefore SDSS J1048+0055 would 
most likely be a galaxy merger with two quasar cores in its central 
region\footnote{Readers may find other lines of evidence that support 
the BBH interpretation in the detailed discussion of the whole sample of 
double-peaked narrow emission line from SDSS DR1 (Wang et al. 2004, in preparation)}.

\subsection{The efficiency of finding BBH using double-peaked narrow
emission lines}

Now that most galaxies harbor a supermassive black hole and the
frequency of galaxy collisions is relatively high, closely bound
BBH in a single galaxy should not be a rare occurrence.
Theoretical calculations show that coalescence of two black hole
goes through several stages. Dynamic friction causes the black
holes along with their surrounding stellar cluster to approach
each other. As the stars in the cluster being stripped off, the
the two black holes form a BBH. Some BBH will further lose angular
momentum through the scattering of passing-by stars, but analysis
shows that some BBH can survive for about one Hubble time in
large, spherical galaxies and the orbital velocities of the
surviving BBH are generally believed to be $\sim 10^{2-4}$ km
s$^{-1}$ (Yu 2002; c.f., Milosavljevi{\' c} \& Merritt 2003).

Detection of supermassive BBH and the estimation of the frequency
of such events can provide important constraints on models of
galaxy formation and evolution. Now, the evolution timescale of
BBH is extremely uneven. A sharp rise occurs at $\sim 10$ pc and
the timescales for a separation of $\lesssim $ a few pc are
several orders of magnitude greater than for $\sim 10^{1-4}$ pc
(Yu 2002). The time it takes a BBH of similar masses to evolve
from a few kpc to $\sim 10$ pc is $\sim 10^{8} $ yr, and this time
increase with increasing mass ratio of the BBH. Therefore, even
every galaxy experiences a major merger during its lifetime and
most of BBH survive, the absolute majority would have separation
$\lesssim $ a few pc, which is beyond the resolution limit of most
of present instruments. While double nuclei with separation of
several kpc or larger have been identified in some galaxies,
however, BBH with separation $\sim 10^{1-3} $pc have not been
found yet.

It is likely that merging will trigger the activity of both black
holes at intermediate separations. Gas concentration and starburst
activity in the nuclei were observed in interacting galaxies at
even large separation (e.g., Gao \& Solomon 2003), and a pair of
binary AGN are detected in NGC 6240. The associated star-cluster
or bulge around each black hole may survive in a fraction of such
systems at this distance. Thus each nucleus may well possess its
own NLR. In this case, our calculations show that about 5-10\% of
such systems will be found as double-peaked sources under the
generic assumption of virialization of NLR. If a significant
fraction of AGN are indeed in such merging systems, a few percent
of AGN should show double-peaked line profile. We found $\sim$ 100
objects that shows this feature in at least one narrow line from
the subsample of $\sim 10^4$ SDSS galaxies and quasars that either
[OIII] or H$\alpha$ have S/N ratio $\gtrsim 30 \sigma $. The
fraction of such objects is about 1\%, and increases with the
level of nuclear activity. Hence, the simple model estimation is
quite consistent with the result of observation. The model and the
sample will be described in detail elsewhere.

\subsection{Other possible explanations and future prospect}

We would like to point out that the BBH interpretation for SDSS
J1048+0055, as well as for the other objects in our double-peaked
narrow line sample, is certainly not the only one. Two other
possible interpretations are bipolar outflow and a disk-like
emission line region. The latter seems unlikely because the
compact structure and flat spectrum of SDSS J1048+0055 suggest
that the system is observed face-on, while an inclined disk is
required to explain the double peaked profile in SDSS J1048+0055.
On the other hand, if the bipolar outflow interpretation is
correct, then we can only explain the weaker radio source as a
one-sided jet like that of 2255-282 which is an optical variable
radio source. For the latter, a jet is revealed by VLBA in the 5
GHz image and its 15 GHz images show radio structure similar to
that of the former (Zensus et al. 2002 and references therein).
The VLBA radio image of SDSS J1048+0055 was taken some 8 years
ago, hence, another high angular resolution radio image can tell
whether this is indeed the case.

About $25\%$ of the objects in our double-peaked narrow line
sample are detected in the radio by the FIRST and $\sim 10$
objects, including SDSS J1048+0055 and SDSS J094144.82+575123.7,
have radio flux $\gtrsim 10$ mJy. According to the estimates made
in \S 3.3, some of these objects should be resolved into two
discrete radio sources at $\sim $ milliarcsec resolution if the
BBH interpretation is indeed correct. A sizable fraction of the
objects in our sample may also be resolved in the optical by HST.

\acknowledgements We thank Prof. T. Kiang for help in English. We are
grateful to Drs Youjun Lu \& Qingjuan Yu for enlightened discussion.
This work is supported
by the Chinese National Science foundation (NSF 19925313 and
10233030) and the key program of Chinese ministry of science and
technology. This paper has made use the data from NED, NRAO, and
SDSS.

Funding for the creation and distribution of the SDSS Archive has been
provided by the Alfred P. Sloan Foundation, the Participating Institutions,
the National Aeronautics and Space Administration, the National Science
Foundation, the U.S. Department of Energy, the Japanese Monbukagakusho, and
the Max Planck Society. The SDSS Web site is http://www.sdss.org/.

The SDSS is managed by the Astrophysical Research Consortium (ARC) for the
Participating Institutions. The Participating Institutions are The
University of Chicago, Fermilab, the Institute for Advanced Study, the Japan
Participation Group, The Johns Hopkins University, Los Alamos National
Laboratory, the Max-Planck-Institute for Astronomy (MPIA), the
Max-Planck-Institute for Astrophysics (MPA), New Mexico State University,
Princeton University, the United States Naval Observatory, and the
University of Washington.

\clearpage

\begin{figure}[tbp]
\epsscale{0.8} \plotone{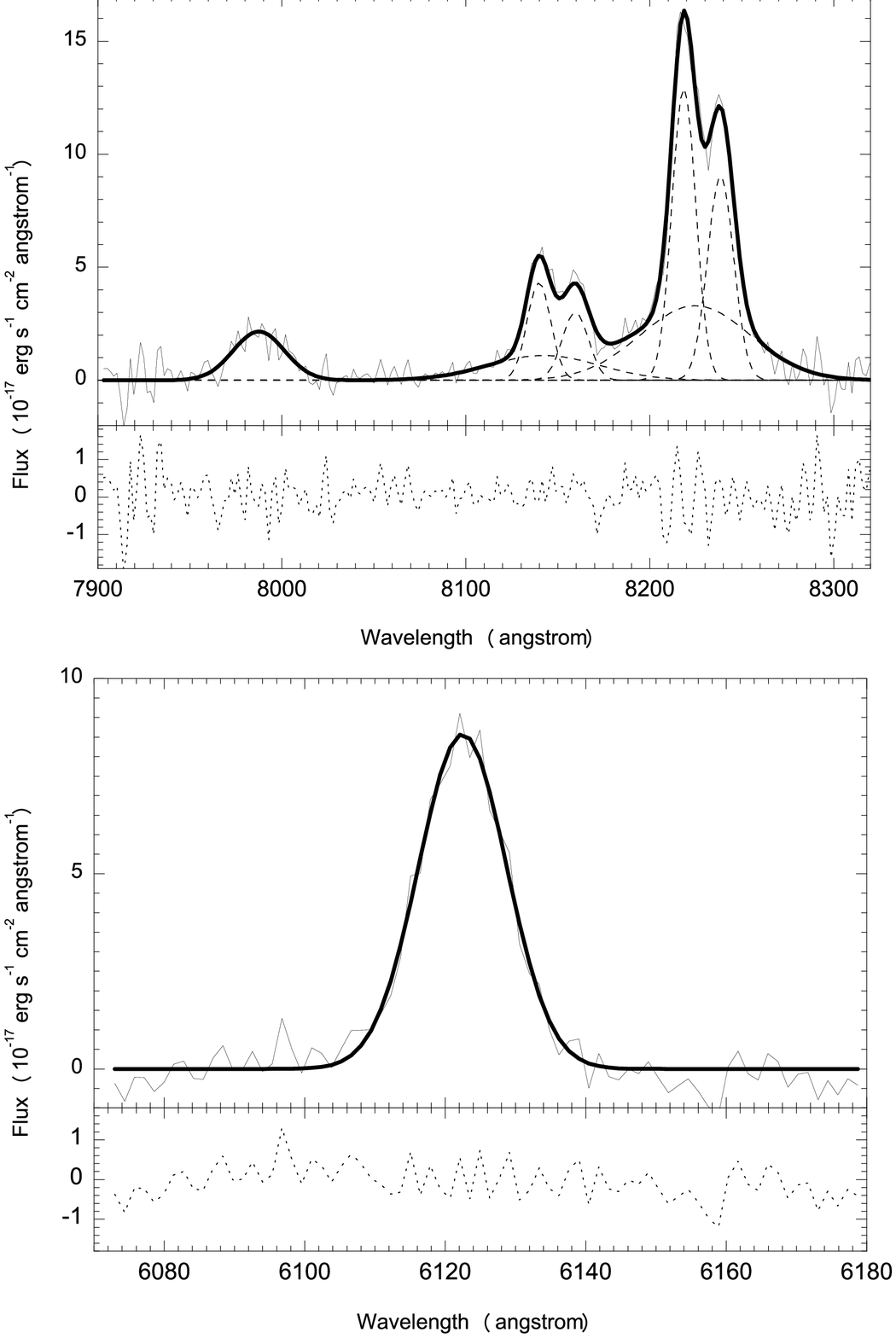} \caption{Gaussian fitting of
H$\protect\beta $+ [OIII] regime (upper panel) and [OII] (lower
panel) after continuum substraction. Dotted lines denote the fit
residuals.} \label{fig-1}
\end{figure}

\begin{figure}[tbp]
\epsscale{1.0}
\plotone{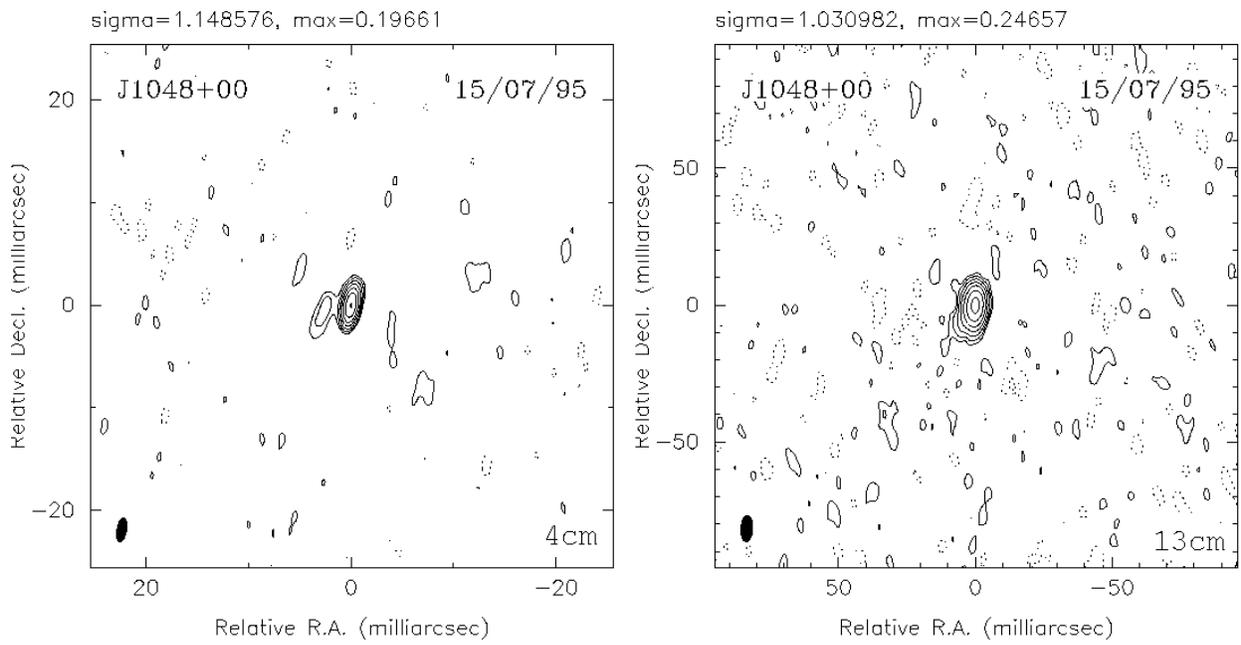} \caption{The 8.4-GHz (left panel)
and 2.3-GHz (right panel) radio morphology of SDSS J1048+0055 from
the VCS1 survey. Contour levels are -1,1,2,4,8, etc. times the
lowest contour level. SDSS J1048+0055 is only marginally resolved
at 2.3-GHz, but higher resolution 8.4-GHz image clearly reveals
two discrete radio sources. }
\label{fig-2}
\end{figure}

\end{document}